\documentclass[12pt]{article}
\usepackage{jheppub}
\usepackage{MnSymbol}

\def\be{\begin{equation}}
\def\ee{\end{equation}}
\def\ba{\begin{aligned}}
\def\ea{\end{aligned}}
\def\ben{\begin{eqnarray}\displaystyle}
\def\een{\end{eqnarray}}

\preprint{
{\small{\textsf{QMUL-PH-11-20}}}}

\title{%A note on the vortex-antivortex 
Factorisation of  $\mathcal{N}=2$ theories on the squashed 3-sphere}

\author[1,2]{Sara Pasquetti}

\affiliation[1]{School of Physics, Queen Mary University of London, \\
Mile End Road, London E1 4NS, UK\\}

\affiliation[2]{Pure Mathematics Section, Huxley Building, \\
Imperial College, Queen's Gate
London SW7 2AZ, UK\\}

\abstract{Partition functions of  $\mathcal{N}=2$ theories on the
squashed 3-sphere  have been recently shown to localise to matrix integrals.
By explicitly   evaluating the  matrix  integral we show that 
abelian partition functions can be expressed  as a sum of products of two blocks.
We identify  the first block with  the  partition function of the vortex theory,
with equivariant parameter $\hbar=2\pi i b^2$, defined on   $\mathbb{R}^2\times S_1$
corresponding to the $b \to 0$ degeneration  of the ellipsoid.
The second block gives  the  partition function of the vortex theory,
with equivariant parameter $\hbar^L=2\pi i/ b^2$,  on the dual  $\mathbb{R}^2\times S_1$
corresponding to the $1/b \to 0$ degeneration.
The ellipsoid partition appears to provide the $\hbar \to \hbar^L$  modular invariant 
non-perturbative completion of the vortex theory.
}    

\begin{document}

\maketitle
\section{Introduction}

Partition functions of  $\mathcal{N}=2$ theories on the
ellipsoid $S^3_b$ have been recently shown to reduce to  matrix integrals  in \cite{hama},
by a generalisation of the  localisation method developed in \cite{pestun,kapo}.
The ellipsoid is a deformation of the three-sphere, preserving  a  $U(1)\times U(1)$ isometry, 
which can be parameterised as:
\be
b^2 |z_1|+\frac{1}{b^{2}}|z_2|^2=1, \qquad z_1,z_2 \in \mathbb{C},
\ee
where $b$ is the squashing parameter.

In the $b\to 0$ limit the  ellipsoid  degenerates to   $\mathbb{R}^2\times S_1$
where the partition function counts finite-energy configurations  on $\mathbb{R}^2$, known as vortices \cite{dgl,DG,DGG}.
The vortex partition function can be evaluated via equivariant localisation, 
with equivariant parameter $\hbar=2\pi i b^2$ \cite{shad}.
More precisely  on $\mathbb{R}^2\times S_1$  we have the so-called K-theory vortex partition function,
with  equivariant parameter  $q=e^{\beta \hbar}$ where $\beta$ is the $S_1$ radius.
The vortex partition function $Z_V$ can be expressed as a perturbative series:
\be
\label{vv}
Z_V=e^{S_0/\hbar+ S_1+ S_2 \hbar +\cdots }.
\ee
On a general basis  we expect   the perturbative series $S_0/\hbar+S_1+ S_2 \hbar+\cdots$ to be  asymptotic 
and as a consequence we expect non-perturbative corrections of the form $e^{-A/\hbar}$.

Clearly we can  realise another completely equivalent construction 
by considering the opposite limit $1/b \to 0$. The ellipsoid degenerates to another  copy of  $\mathbb{R}^2\times S_1$ where lives the dual vortex theory,  that we call antivortex theory $ \bar Z_V$,  with equivariant parameter $\hbar^L=2\pi i/b^2$.
%which can in turn be engineered in topological string and related to open BPS counting.

We  expect that the ellipsoid partition function, in the $b\to 0$ limit,  will have a perturbative part coming from 
the vortex sector and a  non-perturbative  part from the antivortex sector and conversely in the $1/b \to 0$ limit.

As an illustration of what we discussed above let us consider  the double-sine function 
$s_b$ whose definition and relevant identities are collected in appendix \ref{appendix}.
The double-sine is the basic building block appearing in ellipsoid partition functions \cite{hama}
and it is the partition function of a free chiral multiplet.
The double-sine function has the following representation:
\be
\label{ese}
s_b(x)=\frac{e^{-i \pi  x^2/2 }}{\prod_{k=1}^\infty\left(1-q_1^{-(2k+1)}  e^{-2\pi b x}\right)\prod_{k=1}^\infty\left(1-q_2^{-(2k+1)} e^{-2\pi x/b}\right)}\, ,
\ee 
where $q_1= e^{i\pi b^2}$ and $q_2= e^{i\pi/b^2}$.
This expression can be 
interpreted as the factorisation of the ellipsoid partition function into
a vortex and an antivortex  contribution.
By observing the   ``free energy'':
\ben
\log s_b(x)&=&\frac{-i \pi x^2}{2}
+\left( \sum_k \frac{1}{k} \frac{e^{-2\pi x b k}  }{(q_1^k-q_1^{-k})}\right)+
\left( \sum_k \frac{1}{k}  \frac{e^{-2\pi  x/ b k}  }{(q_2^k-q_2^{-k})}\right)\, ,
\een
we see that  the small $\hbar$ expansion produces 
a perturbative part in $\hbar$, from the vortex sector,  and a  non-perturbative part in  $e^{-1/\hbar}$ from the antivortex sector.
A similar structure appears in the dual  small $\hbar^L$ expansion.

On this respect the ellipsoid partition function  appears  as the non-perturbative completion of the vortex theory.
It is interesting to observe that the inclusion of non-perturbative corrections restores  the modular $b \to 1/b$ invariance
which is indeed manifest in the ellipsoid\footnote{For a similar discussion in the context of matrix models see \cite{nonp}.}.

In this note we will argue that the vortex-antivortex factorisation occurs  also in interacting cases
where the partition function is a matrix integral. 
By a direct evaluation of the integral  we show that
the partition function can be expressed as:
 \ben
 \label{main}
Z_{S^3_b}= \sum_{i=1}^N  Z^{(i)}_{cl}
Z^{(i)}_{1-loop} Z^{(i)}_{V}\times \bar Z^{(i)}_{1-loop} \bar Z^{(i)}_{V},
\een
where, as we will  discuss in the next section, the sum runs over vortex sectors.
 
This expression is  reminiscent of the  partition function on the 
4-sphere obtained by Pestun \cite{pestun}, with vortices playing here the role
of  4d instantons.
This analogy  suggests that it should be possible to derive the factorised form of the ellipsoid partition function
directly by an alternative  localisation of the path integral.

The vortex partition function can be geometrically engineered \cite{dgl,btz},
in particular, as shown  in the next section, the vortex partition function is given by an   open topological string partition function
 $Z_V=Z_{\rm top}$.
From this perspective our expression (\ref{main})
is also evocative  of the  Ooguri-Strominger-Vafa relation between  black hole entropy and  topological
string partition function \cite{osv}:
$$
Z_{\rm BH}= |Z_{\rm top}|^2.
$$

This note is organised as follows.
In the next section we  study the abelian theory with  $N$ chirals and $N$ antichirals,
beginning  with the explicit evaluation of the matrix integral that leads to the 
factorised form.
We then  interpret  our expression in terms of vortex theories and topological strings.
In section \ref{s2} we repeat the analysis for  the  abelian  theory with $2N$ chirals.
In section \ref{s3} we discuss more general theories and future directions.

\section{The non-chiral theory}\label{s1}

Our first example is  the $\mathcal{N}=2$
$U(1)$ theory with $N$ flavours.
% ($N$ chiral  with positive charge and and $N$ chirals with negative charge). 
We introduce  vector masses  $m_i$  with $\sum_i^N m_i=0$,
axial masses $\mu_i$, $i=1\cdots N$ and a FI parameter $\xi$.
%In fact we have $N$ vector and $N$ vector masses but we can shift one parameter away.
The ellipsoid partition function, as explained in \cite{hama}, reduces to a matrix integral
which reads:
\ben
\label{nc}
Z^{U(1)}_{S_3^b}(N,N)=\int dx\, e^{2 \pi i x \xi}\prod_{i=1}^N \frac{s_b(x+ m_i +\mu_i/2 + i Q/2)   }{s_b(x+ m_i -\mu_i/2 - i Q/2)   },
\een
where $Q=b+1/b$. It is possible to   directly evaluate the matrix integral 
(\ref{nc}) by closing  the integration contour in the upper half-plane and picking the contributions of the  simple 
poles of the double-sines in the numerators located at:
  \ben
  \label{poles}
x=-m_i-\mu_i/2 + i \,m\, b+i \,n/b, \qquad m,n\geq0, \qquad i=1\cdots N.
\een
There are  $N$ infinite towers of poles to take into account.
Each tower contains poles labelled by  indices  $m,n$ and  the residue  at each $(m,n)$ pole
 can  be evaluate by  means of  formulas \ref{master} and \ref{master2}.
Putting all together we find:
 \ben
 \label{facto}
Z^{U(1)}_{S^3_b}(N,N)= \sum_{i=1}^N\frac{e^{-2\pi i \xi  (m_i+\mu_i/2) }}{s_b( C_{ii}-iQ/2)}\prod_{j\neq i}^N \frac{s_b(D_{ji}+i Q/2)}{ s_b(C_{ji}-i Q/2)}
Z^{(i)}\bar Z^{(i)}\, ,
\een
 where: 
  
  \ben
  \label{hyper}
 Z^{(i)}
 %&\sum_{n=0}^\infty \prod_{k=1}^{n}
   % \prod_{j=1}^N
 %\frac{
  %(1-q_1^{-2(k-1)}  e^{- 2\pi b C_{ji}})}{ 
% (1-q_1^{-2k} e^{-2\pi b D_{ji}}) }   \left( e^{-\pi b  \sum_{j=1}^N(\mu_j+i Q)} e^{-2\pi \xi b  }\right)^{n}\nonumber=\\ 
 &=&\sum_{n=0}^\infty \prod_{k=1}^{n}
  \frac{   (1-q^{(k-1)}  e^{-2\pi b C_{ii}})}{  (1-q^{k} )}
 \prod_{j\neq i}^N
 \frac{  (1-q^{(k-1)}  e^{- 2\pi b  C_{ji}})}{  (1-q^{k} e^{-2\pi b D_{ji}}) }  z^n\equiv Z^{(i)}_{V}\, ,
 \een
 \ben
\label{bhyper}
 \bar Z^{(i)}=
 %&=&\sum_{m=0}^\infty \prod_{l=1}^{m}
  %\prod_{j=1}^N
 %\frac{(1-q_2^{-2(l-1)}  e^{- 2\pi/b C_{ji}})}{ 
 %(1-q_2^{-2l} e^{-2\pi/ b D_{ji}}) }   \left( e^{-\pi/b  \sum_{j=1}^N ( \mu_j+iQ)} e^{-2\pi \xi /b  }\right)^{m} =\nonumber\\ 
 %&=&
 \sum_{m=0}^\infty \prod_{l=1}^{m}
   \frac{   (1-\bar q^{(l-1)}  e^{-2\pi/b  C_{ii}})}{  (1-\bar q^{l} )}
  \prod_{j\neq i}^N
 \frac{
  (1-\bar q^{(l-1)}  e^{- 2\pi/b  C_{ji}})}{
 (1-\bar q^{l} e^{-2\pi/ b D_{ji}}) }  \bar z^m\equiv\bar Z^{(i)}_{V}\, ,
   \een
with:
\ben
\label{dec}
 && D_{j i}= m_j+\mu_j/2-m_i-\mu_i/2, \qquad  C_{j i}=m_j-\mu_j/2-m_i-\mu_i/2\, .
 \een
We also introduced equivariant parameters:
\ben
\label{para1}
q=q_1^{-2}=e^{-\beta \hbar}, \qquad  \bar q=q_2^{-2}=e^{-\beta \hbar^L}\, ,
\een
where $\beta$ is the $S^1$ radius in $\mathbb{R}^2\times S^1$
and  vortex counting parameters:
\ben
\label{para2}
 \left(e^{-\pi b  \sum_{j=1}^N  (\mu_j+iQ)} e^{-2\pi \xi b  }\right)= e^{2\pi i b\xi_{\rm eff}} =z, \quad  \left(e^{-\pi/ b  \sum_{j=1}^N ( \mu_j+iQ)} e^{-2\pi \xi /b  }\right)=e^{2\pi i \xi_{\rm eff}/b} =\bar z\, ,
\een
with $\xi_{\rm eff}$ the effective FI parameter.

The factorisation  of the matrix integral (\ref{facto})
at a merely technical level occurs because of   the cancellation of the factors $(-1)^{n m}$
in formulas (\ref{master}), (\ref{master2}) between double-sine functions in the numerator 
(chiral) and in the denominator (anti-chirals).
For the same reason an integral involving an arbitrary number of chirals
will factorise whenever the difference between the number of chirals and anti-chirals is  even.
This condition is however necessarily satisfied by physical partition functions.
% it is  just the condition for the cancelation of the $Z_2$ anomaly.
Indeed a one loop computation  yields the following expression for the effective Chern-Simons coupling,
obtained integrating out  charge $Q[\psi]$ fermions:
\be
k_{\rm eff}=k+\sum_{\rm fermions} Q[\psi]^2/2.
\ee
A necessary condition to ensure gauge invariance is to have $k_{\rm eff}\in \mathbb Z$ which is precisely the 
condition for the factorisation of the  partition function.
%This condition, for theories without a bare Chern-Simons coupling, is precisely our  condition for the factorisation of the integral.
%\footnote{
%We observe here that the expansion parameter is the effective FI parameter:
%\be
%\xi_{eff}=\xi+\frac{1}{2}\sum_{j=1}^N ( \mu_j+iQ).
%\ee
%due to a mixed cs \cite{dt}.}

We could have included  in the action  a level $k$  Chern-Simons term,
which, after localisation, contributes to the integrand as  $e^{-\pi i k x^2}$  \cite{kapo}. 
When evaluated at the poles (\ref{poles}) the Chern-Simons term gives:
\ben
\label{cs}
\nonumber e^{-\pi i k (-m_i- \mu_i/2+i m b +i n/b)^2}=
e^{-\pi i k (m_i+ \mu_i/2)^2 }e^{-2\pi k (m_i+ \mu_i/2)(mb+n/b) }  q_1^{k m^2}  q_2^{k n^2} (-1)^{ k n m}\, .
 %=\\=
 %e^{-\pi i k (m_i+ \mu_i/2)^2 }  e^{-2\pi k (m_i+ \mu_i/2)(mb+n/b) }  q^{- (k m^2/2)}  \bar q^{- (k n^2/2)}  (-1)^{2 \pi i k n m}\, .\nonumber
 \\
 \een
Clearly the inclusion  of a Chern-Simons term with $k\in \mathbb Z$ in our theory (\ref{nc})  will not spoil the factorisation property.
It is also easy to realise that for theories where the difference between the number of chirals and anti-chirals is  odd,
a bare  Chern-Simons term with $k\in \mathbb{Z}/2$  will ensure the factorisation.

We close this section by observing that  $Z^{(i)}_V$ is just a  basic hypergeometric series:
 \be
  Z^{(i)}_V={}_{N}\Phi^{(i)}_{N-1}(a^{(i)}_{1},\mathellipsis,a^{(i)}_{N};b^{(i)}_{1},\mathellipsis,b^{(i)}_{N-1};z,q)=
  \sum_n \frac{\prod_{j=1}^N (a^{(i)}_j,q)_n}{(q,q)_n\prod_{j\neq i}^N (b^{(i)}_j,q)_n} z^n \, ,
 \ee
 where:
 \ben
 a^{(i)}_j=e^{-2\pi b C_{ji}} , \qquad b^{(i)}_j=q e^{-2\pi b D_{ji}}\, ,
  \een
 and the q-Pochhamers are defined as:
\be
(a,q)_n=\prod_{k=0}^{n-1}(1-a q^k).
\ee
In particular, by looking at the definition of  coefficients $C_{ji},D_{ji}$ given in (\ref{dec}),
it  is  easy to realise that the index  $i=1,\cdots N$ labels  the $N$  independent solutions of the basic-hypergeometric difference equation.
Analogously  $\bar Z^{(i)}_V$ for $i=1,\cdots N$ will be  basic hypergeometric series
defined as above with  $b\to 1/b$ .
%\ben
%\label{fde}
%\left[\prod_{j\neq i}^N\left(1-b^{(i)}_j q^{-\partial_u-1}\right)\left(1- q^{-\partial_u}\right)-z \prod_j^N \left(1-a^{(i)}_j q^{-\partial_u}\right) \right]Z^{U(1)}_{S^3_b}(N,N)=0
%\een
%with $z=e^u$ and $i=1\cdots N$.
In conclusion we find that the  ellipsoid partition function will  be annihilated by a $q$-difference and by a $\bar q$-difference operator.
It would be interesting to explore the role of these operators in the dual Chern-Simons theory as discussed in \cite{DGG}.

\subsection{Vortex-Antivortex factorisation}\label{ss1}

As we mentioned in the introduction, in the $b\to 0$
limit the ellipsoid degenerates to  $\mathbb{R}^2\times S^1$,
therefore, in this limit, we expect the ellipsoid partition function to reduce to the vortex theory on $\mathbb{R}^2\times S^1$
\cite{dgl, DG,sv}.

The vortex  partition function  can be computed via equivariant localisation  and consists of  a perturbative part and a vortex part \cite{shad}.
Since we are considering abelian theories, we are interested 
in vortex configurations   labelled by the trivial partition  $1^n$.
The (K-theory) vortex partition function for the abelian theory with $N_f$ fundamentals and $N_a$ anti-fundamentals reads:
\be
%Z_V=
\sum_n z^n  \frac{  \prod_{i=1}^{N_f}\prod_{k=1}^n(1-Q^f_i q^{k-1})}{\prod_{k=1}^n(1-q^k)  \prod_{i=1}^{N_a}\prod_{k}^n(1-Q^a_i q^{k})},
\ee
where $Q^f_i=e^{-\beta m^f_i}$, $Q^a_i=e^{-\beta m^a_i}$
and $m^{f}_i$, $m^{a}_i$  are  fundamental and anti-fundamental masses.
In the $\beta \to 0$ limit we recover the (homological) vortex partition function on $\mathbb{R}^2$ computed in \cite{shad}.

It is immediate to identify our expressions  $Z^{(i)}_V$, for $i=1,\cdots N$, as 
abelian vortex partition functions  with $N$ chiral fundamental multiplets with masses $D_{ji}$ for $j\neq i$ (one is always massless)  and $N$ chiral antifundamentals with masses $C_{ji}$.

Similarly  $\bar Z^{(i)}_{V}$, for $i=1,\cdots N$, can be identified with 
dual ($\hbar \to \hbar^L$)  vortex partition functions with $N$ fundamental and $N$ anti-fundamental chirals
on the dual  $\mathbb{R}^2\times S^1$ corresponding to the $1/b \to 0$ degeneration  of the ellipsoid.

We now move to the study of the prefactor  in eq (\ref{facto}).
By using the the representation (\ref{ss}) it is easy to see that the prefactor  is also factorised in terms of $1-$loop contributions given in \cite{shad}:
\ben
\label{1l}
 \frac{ \prod_{j\neq i}^N  s_b(D_{ji}+i Q/2)}{\prod_{j}^N  s_b(C_{ji}-i Q/2)}&\equiv&
e^{\pi i/2\sum_{j}\left( (D_{ji}+iQ/2)^2- (C_{ji}-iQ/2)^2 \right)}  Z^{(i)}_{1-loop} \bar Z^{(i)}_{1-loop}\, ,\een
with
\ben
Z^{(i)}_{1-loop} =\prod_{k=1}^\infty
\frac{\prod_{j\neq i}^N \left(1-q^k e^{-2\pi b D_{ji}}  \right)}{\prod_{j}^N  \left(1-q^{k-1} e^{-2\pi b C_{ji}}  \right)},
\qquad \bar 
Z^{(i)}_{1-loop} =\prod_{k=1}^\infty
\frac{\prod_{j\neq i}^N \left(1-\bar q^k e^{-2\pi/b D_{ji}}  \right)}{\prod_{j}^N  \left(1-\bar q^{k-1} e^{-2\pi/b C_{ji}}  \right)}.
\een
Finally the  exponential terms in eq. (\ref{1l}) and in eq. (\ref{facto}) 
can be simplified and combined to give the $i-$th classical action:
\be
e^{-2 \pi i \xi (m_i+\mu_i/2)}e^{i\pi/2\left( (D_{ji}+iQ/2)^2- (C_{ji}-iQ/2)^2 \right)}=e^{-\pi i \sum_j \mu_j m_j} e^{-2\pi i \xi_{\rm eff}  (m_i+\mu_i/2) } \equiv e^{-\pi i \sum_j \mu_j m_j} Z^{(i)}_{cl}\,,
\ee
where $\xi_{\rm eff}$  was introduced in (\ref{para2}).

%In particular the leading part of $Z^{pert}\sim e^{\mathca{W}^{pert}/\hbar}$ gives the one-loop part of the twisted superpotential:

%\be\mathcal{W}^{pert}=\sum_j^{N_f} (x+m^f_i)\left(\log|\frac{x+m^f_i}{\Lambda}|-1\right)-\sum_j^{N_a} (x+m^a_i)\left(\log|\frac{x+m^a_i}{\Lambda}|-1\right)
%\ee

To summarise, up to a prefactor, we can write the ellipsoid partition function in the following form:
 \ben
 \label{va}
Z^{U(1)}_{S^3_b}(N,N)= \sum_{i=1}^N  Z^{(i)}_{cl}\times 
\left(Z^{(i)}_{1-loop} Z^{(i)}_{V} \right) \times  \left(\bar Z^{(i)}_{1-loop} \bar Z^{(i)}_{V}\right)\, ,
\een
which makes visible  the vortex-antivortex structure  for finite $b$.

As we explained in the introduction,
the semiclassical $\hbar$ expansion of  the  free energy
will consist of a perturbative $\hbar$ part  coming from the vortex sector  in (\ref{va}) and a non-perturbative part
coming from the anti-vortex sector.
Conversely, in the dual semiclassical $\hbar^L$ expansion, the anti-vortex sector  will contribute
perturbatively  while the vortex sector non-perturbatively.

We can then regard the ellipsoid partition function as  the 
modular   $\hbar \leftrightarrow \hbar^L$ invariant
non-perturbative completion of the vortex theory.

Our final expression for the ellipsoid partition function (\ref{va}) has the same structure as the partition function of 
 $\mathcal{N}=2$ theories on the four sphere $S^4$ obtained  by Pestun \cite{pestun}, which reads:\footnote{We thank G.~Bonelli and N.~Drukker for discussion on this point.}
\be
\label{pes}
Z_{S^4}=\int d\mu_\alpha Z_{cl}(\alpha) |Z_{1-loop}(\alpha)|^2 |Z_{inst}(\alpha,\tau)|^2\, ,
\ee
where  $d\mu_\alpha$ is the measure over  the Cartan of the gauge group.
In the ellipsoid case (\ref{va}) the role of  the instanton partition function $Z_{inst}(\alpha, \tau)$ is played by the vortex partition function, with instanton and vortex counting parameters 
given respectively  by the 4d gauge coupling and by the 3d FI.

However while the $S^4 $ partition function
(\ref{pes})  factorises into  holomorphic and antiholomorphic blocks,
  the ellipsoid partition function factorises into two blocks  related by $\hbar \to \hbar^L$.
Moreover  while in the $S^4$ cases there is an integration 
over  the Cartan of the gauge group, in the ellipsoid case (\ref{va}) we have a discrete sum.

It should be possible to obtain  our result (\ref{va})
directly from the localisation of the path integral by choosing an appropriate Q-exact term to add to the action.

\subsection{Geometric engineering}\label{ss3}

\begin{figure}[!ht]
\leavevmode
\begin{center}
\includegraphics[height=6cm]{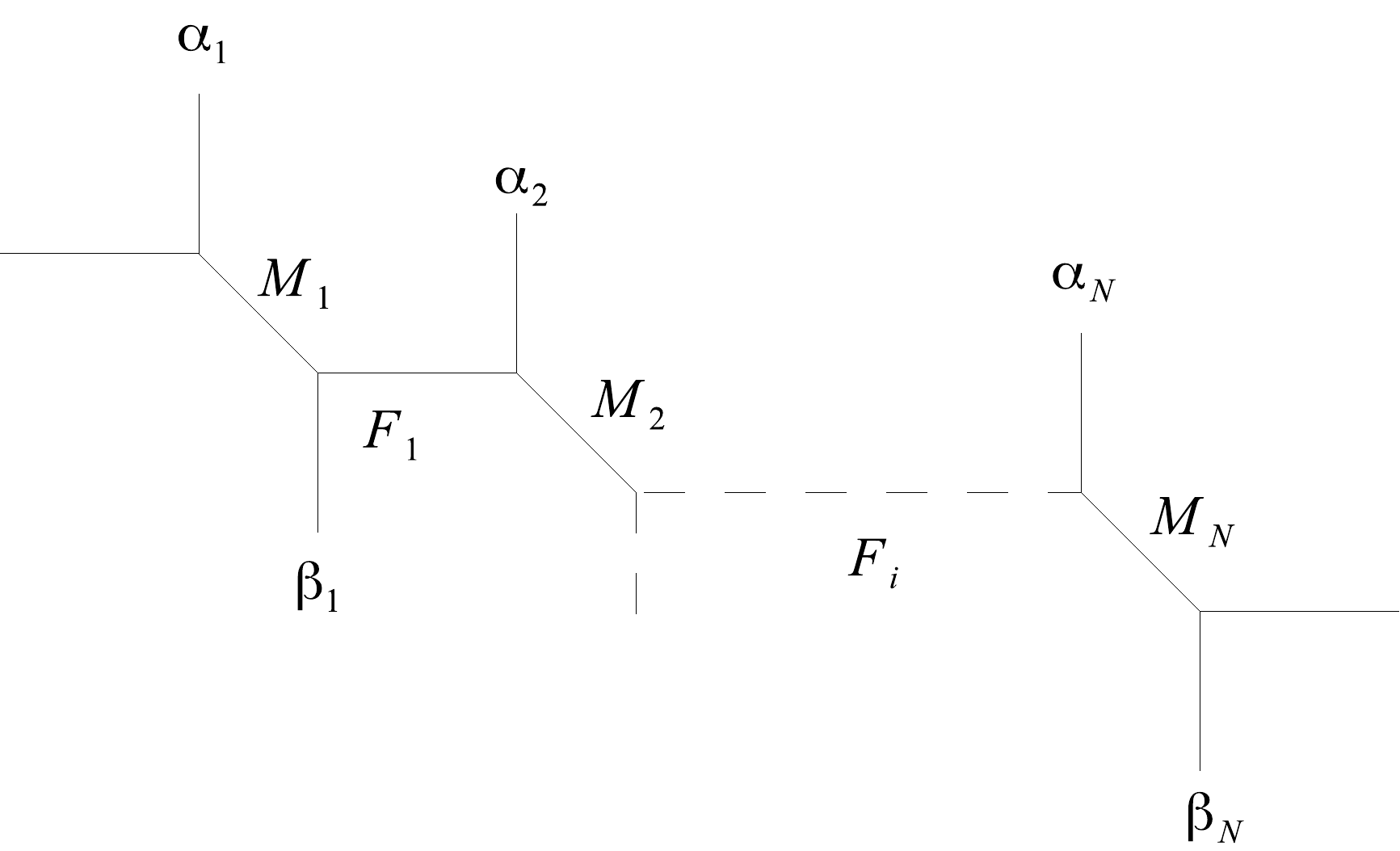}
\end{center}
\caption{The strip geometry.}
\label{bstrip}
\end{figure}

As  explained in \cite{dgl} (see also \cite{btz}) the K-theory vortex partition function can be  
engineered in topological string theory.
Here we observe  that the relevant topological string geometry arises very naturally directly from 
the semiclassical analysis of the ellipsoid  partition function.
We start from the integral form of the partition function (\ref{nc}), shift the integration variable 
$x\to x- m_i-\mu_i/2+i Q/2$ and  use the $b \to 0$  limit of the double-sine function given in (\ref{limit}) to obtain:
\ben
\label{sa}
Z^{U(1)}_{S^3_b}(N,N)&\sim&\int dx\,
e^{2\pi i x \xi}\, e^{-\frac{\pi i}{2}\sum_j \left(
 (x+D_{ji} +i Q)^2-(x+C_{ji})^2
\right)}\nonumber  \\&& \times \, e^{\frac{1}{2\pi i b^2}\sum_j\left[
{\rm Li}_2(-e^{2\pi b (x+D_{ji} +i Q)})-
{\rm Li}_2(-e^{2\pi b (x+C_{ji} )}) \right]}\sim \int dx\, e^{\mathcal{W}(x)}\, ,
\een
where $\mathcal{W}(x)$ is the twisted superpotential.
By extremising  we find:
\be
\label{ex}
0=\partial_x\mathcal{W}(x)=2\pi \xi -\pi i \sum_{j} (D_{ji} -C_{ji}+iQ)
-\frac{1}{i b}\sum_j \left(
\log(1+e^{2\pi b (x+D_{ji} +i Q)})-
\log(1+e^{2\pi b (x+C_{ji} )}) \right).
\ee
If now we introduce  $\mathbb{C}^*$ coordinates
$X\equiv e^{2\pi b x}$ and $Y\equiv e^{2\pi i b \xi_{\rm eff}} $, we can rewrite the  saddle point equation (\ref{ex}) as:
\be
\label{mirror}
Y=\prod_j \frac{(1+ X e^{2\pi b (D_{ji} +i Q)}) }{(1+X e^{2\pi b C_{ji}  })}\,,
\ee
which can be immediately identified with the the mirror curve  {\it \`a la} Hori-Vafa \cite{hv} of the  strip geometry depicted  in
Fig. \ref{bstrip}.

The leading contribution to the integral (\ref{sa}), the twisted superpotential $\mathcal{W}(x)$,
is the Abel-Jacobi map on the mirror curve,
or in other words,  it is the disk amplitude with boundary  on the mirror of  a toric brane
\cite{AV,AKV}.
At this point we could use the Remodelling  method \cite{remo} to reproduce all the subleading terms in the semiclassical expansion 
as in \cite{ab}.
Equivalently we can use  the topological vertex \cite{vertex} to  compute A model 
amplitudes directly reproducing the vortex counting partition function as in \cite{dgl},\cite{btz}.
We  choose the second approach.

We need to compute  open topological $A$ model partition functions
on the strip geometry   with boundaries mapped into a single  toric brane placed in one
of the {\it gauge legs}.

The open strip partition function  is given by \cite{strip}:
\ben
\frac{\mathcal{K}^{\alpha_1\cdots \alpha_N}_{\beta_1\cdots \beta_N} }{
\mathcal{K}^{\bullet \cdots \bullet}_{\bullet\cdots \bullet}}=
\left(\prod_l^N s_{\alpha_l}(q^\rho) s_{\beta_l}(q^\rho)\right)
\prod_k\frac{\prod_{i\leq j} (1-q^k Q_{\alpha_i\beta_j})^{C_k(\alpha_i,\beta_j^t)}    \prod_{i< j} (1-q^k Q_{\beta_i\alpha_j})^{C_k(\beta_i^t \alpha^t_j)}}{   \prod_{i< j} (1-q^k Q_{\alpha_i\alpha_j})^{C_k(\alpha_i,\alpha_j^t)} \prod_{i< j} (1-q^k Q_{\beta_i\beta_j})^{C_k(\beta_i^t,\beta_j) }}\, ,\nonumber\\
\een
we are interested in  the case where the $i-$th representations is equal to a column representation $\alpha_i=1^n$ and all the other reps are trivial. 
By using that:
\ben
\label{ck}\nonumber
C_k(1^n,\bullet)=1\qquad &{\rm for}& \qquad k\in[0,n-1],\\
 C_k(n,\bullet)=1  \qquad   &{\rm for}& \qquad k\in[-n+1,0]\, ,
\een
 and otherwise zero we find:

\ben
\label{strip}
A^{(i)}_n\equiv\frac{\mathcal{K}^{\bullet\cdots1^n\cdots \bullet}_{\bullet\cdots \bullet} }{
\mathcal{K}^{\bullet \cdots \bullet}_{\bullet\cdots \bullet}}= \frac{1}{\prod_{k=1}^{n}(1-q^k)} \frac{\prod_{j\geq i} \prod_{k=1}^{n} (1-Q_{\alpha_i\beta_j} q^{k-1})     \prod_{j< i} \prod_{k=1}^{n} (1-Q_{\beta_i \alpha_j} q^{-(k-1)})       }{
  \prod_{j>i}\prod_{k=1}^{n} (1-Q_{\alpha_i \alpha_j} q^{k-1})  \prod_{j< i}\prod_{k=1}^{n} (1-Q_{\alpha_i \alpha_j} q^{-(k-1)}) }\, ,\nonumber\\
\een
where the K\"ahler parameter are defined by:
\ben
Q_{\alpha_i\alpha_j}&=&\prod_{k=i}^{j-1} M_k F_k\, , \nonumber\\
Q_{\alpha_i\beta_j}&=&Q_{\alpha_i\alpha_j} M_j \, ,\nonumber\\
Q_{\beta_i\alpha_j}&=&Q_{\alpha_i\alpha_j} M_i^{-1}\, , \nonumber\\
%Q_{\beta_i\beta_j}&=&Q_{\alpha_i\alpha_j} M_i^{-1} M_j \nonumber\\
\een
with $i<j$.
We then construct the generating function:
\be
Z^{(i)}_{top}=\sum_n A^{(i)}_n z^n,
\ee
which can be immediately identified with  the vortex partition function $Z^{(i)}_V$,  given in (\ref{hyper}),
with the  following dictionary:
\ben
\label{dictio}
&& Q_{\alpha_i\alpha_j}=q e^{-2\pi b D_{ij}}, \qquad Q_{\alpha_i\alpha_j}^{-1}=q e^{-2\pi b D_{ji}} \qquad  \nonumber \\
&&  Q_{\alpha_i\beta_j}=e^{-2\pi b C_{ij}},\qquad  Q_{\beta_i \alpha_j}^{-1}= e^{-2\pi b C_{ji}}\, ,
\een
together with a shift of the open modulus $z\to M_{i-1}^{-1} z$.
It is also easy to realise that the  distinct  vortex partition functions $Z^{(i)}_V$ labelled by the index $i=1,\cdots, N$ correspond to open topological string amplitudes with the toric brane placed in 
the $i-$th  {\it gauge leg}.

So far in our computations we have assumed   canonical framing for the toric brane in the representation $\alpha_i$.
To allow $k$ units of framing we need to multiply the strip amplitude (\ref{strip}) by the following term\footnote{
$
\kappa(\lambda)=\sum \lambda_i (\lambda_i -2i +1)=n-n^2
$ for $\lambda=1^n$.}:

\be
(-1)^{k\lambda} q^{k\, \kappa(\lambda)/2}=
(-1)^{kn} q^{-k(n^2-n)/2}.
\ee
By comparing the above expression with eq. (\ref{cs})
we realise that  the inclusion of  a level $k$ Chern-Simons term in the ellipsoid partition function 
amounts to the addition of  $k$ units framing for the brane together with a redefinition 
of the open modulus\footnote{In the context of  geometric engineering of  instanton partition functions 
the level of the 5-dimensional Chern-Simons term has also been related to  framing \cite{tac}.}.
%This latter extra shift from a field theory viewpoint is a one-loop effect, where the FI is shifted by a bare CS.

%We conclude this section by noticing that could have computed the open topological partition function  by means of the refined topological vertex \cite{ref}.
%In this case to  reproduce the vortex partition function we need to take the
% $\epsilon_1=\epsilon,\epsilon_2\to 0$ limit \cite{dgl}.
%However we also have to slightly  modify the dictionary (\ref{dictio}) to map mass parameters
%to Ka\"hler paramters  with  extra shifts of $\epsilon_1,\epsilon_2$.
%In particular for each of the $N$ vortex partition functions we need to adjust the  $\epsilon_1,\epsilon_2$ shift in the dictionary.

\section{The chiral theory}\label{s2}

In this section we  study the  $\mathcal{N}=2$  $U(1)$ theory with $2N$ chirals with  charge $+1$.
%Since we are considering an even number of chirals we do not need to add bare Chern-Simons half-units 
 %to cancel the $\mathbb{Z}_2$ anomaly.
We introduce axial masses $\mu_i$   with $\sum_i^{2N}\mu_i=0$ and FI parameter $\xi$.
The partition function reads:
\ben
\label{c}
Z^{U(1)}_{S^3_b}(2N,0)=\int dx\, e^{2 \pi i \, x\, \xi}\prod_{j=1}^{2N} s_b(x+\mu_j/2 + i Q/2)  \,  .
\een
As in the previous section to evaluate the integral (\ref{c}) we close the contour in the upper half-plane and pick the contributions 
of the poles located at:
  \ben
  \label{spoles}
x=-\mu_i/2 + i\, m \,b+i\, n/b, \qquad m,n\geq 0\, .
\een
Thanks to repeated applications of  eq. (\ref{master}) it is possible to express the partition function in the following  factorised form:
 \ben
 \label{cfacto}
Z^{U(1)}_{S^3_b}(2N,0)= \sum_{i=1}^{2N}
e^{-\pi i \, \xi \,  \mu_i }\prod_{j\neq i}^{2N} s_b(E_{ji}+i Q/2)
Z^{(i)}\bar Z^{(i)}\, ,
\een
 where we introduced  $ E_{j i}=\mu_j/2-\mu_i/2$ 
 and defined:
 \ben
  \label{ch}
 Z^{(i)}= \sum_{n=0}^\infty
  \frac{\left(q^{ n(n+1)/2} (-1)^{n }\right)^{N}}{ \prod_{k=1}^n(1-q^k)} 
 \frac{e^{-2\pi b n \left(\xi- N  \mu_i/2\right) } }{    \prod_{j\neq i}^{2N}
\prod_{k=1}^{n}(1-q^{k} e^{-2\pi b E_{ji}}) }    \, ,\een
 and 
\ben
  \label{bch}
 \bar Z^{(i)}=
 \sum_{m=0}^\infty
  \frac{\left(\bar q^{ m(m+1)/2} (-1)^{m }\right)^{N}}{ \prod_{k=1}^m(1-\bar q^k)} 
 \frac{e^{-2\pi m /b \left(\xi- N  \mu_i/2\right) } }{     \prod_{j\neq i}^{2N}
\prod_{k=1}^{m}(1-\bar q^{k} e^{-2\pi/ b E_{ji}}) } \,.
  \een
We can include a Chern-Simons term in the partition function (\ref{c}) which, when evaluated at the poles (\ref{spoles}) will contribute as:
\be
\label{cs2}
e^{-\pi i k(\mu_i/2+i m b + i n /b)^2 }\, .
\ee
In particular a Chern-Simons term  with  level $k=N$
simplifies  our expressions since  
it cancels  the  factors $\left( q^{ n(n+1)/2} (-1)^{n }\right)^{N}$  in (\ref{ch})  and $\left( \bar q^{ m(m+1)/2} (-1)^{m }\right)^{N} $
in (\ref{bch}) and  shifts the FI parameter. So in this case  we obtain:
\ben
\label{cvo}
  Z^{(i)}_{k=N}= \sum_{n=0}^\infty
  \frac{1}{ \prod_{k=1}^n(1-q^k)} 
 \frac{z^n }{    \prod_{j\neq i}^{2N}
\prod_{k=1}^{n}(1-q^{k} e^{-2\pi b E_{ji}}) } \equiv   Z^{(i)}_V\, , \een
and
\ben
  \bar Z^{(i)}_{k=N}= \sum_{m=0}^\infty
  \frac{1}{ \prod_{k=1}^m(1-\bar q^m)} 
 \frac{z^n }{    \prod_{j\neq i}^{2N}
\prod_{k=1}^{n}(1-\bar q^{k} e^{-2\pi/ b E_{ji}}) } \equiv   \bar Z^{(i)}_V\, , \een
with
\be
z_i \equiv e^{-2\pi b (\xi+i N Q/2)}=e^{2\pi i b \xi_{\rm eff }}\, , \qquad \bar z_i \equiv e^{-2\pi/b(\xi +i N Q/2) }=e^{2\pi i /b \xi_{\rm eff }}\, .\ee

As expected $ Z^{(i)}_V$,  $ \bar Z^{(i)}_V$ are identified with 
K-theory abelian  vortex partition functions with $2N$  chirals.

We now use eq. (\ref{ss}) to 
rewrite  the prefactor  in eq (\ref{cfacto}) in terms of  one-loop factors:
\ben
\label{c1l}
 \prod_{j\neq i}^{2N}  s_b(E_{ji}+i Q/2)=
e^{\pi i/2\sum_{j\neq i}(E_{ji}+iQ/2)^2}  Z^{(i)}_{1-loop} \bar Z^{(i)}_{1-loop}\, ,\een
with
\ben
\label{con}
Z^{(i)}_{1-loop} =\prod_{k=1}^\infty
\prod_{j\neq i}  \left(1-q^{k} e^{-2\pi b E_{ji}}  \right)\, ,
\qquad \bar 
Z^{(i)}_{1-loop} =\prod_{k=1}^\infty
\prod_{j\neq i} \left(1-\bar q^{k} e^{-2\pi E_{ji}/b}  \right)\, .
\een
Finally we introduce the $i-$th classical action:
\be
Z^{(i)}_{cl} = e^{-\pi i \xi_{\rm eff}  \mu_i}\, .
\ee
Putting all together 
the ellipsoid partition function of the $U(1)$ theory with $2N$ chirals 
and level $k=N$ Chern-Simons   can be expressed (up to a prefactor) as:
\ben
 \label{cva}
Z^{U(1),k=N}_{S^3_b}(2N,0)= \sum_{i=1}^N  Z^{(i)}_{cl}\times 
\left(Z^{(i)}_{1-loop} Z^{(i)}_{V} \right) \times  \left(\bar Z^{(i)}_{1-loop} \bar Z^{(i)}_{V}\right)\, .
\een

\begin{figure}[!ht]
\leavevmode
\begin{center}
\includegraphics[height=6cm]{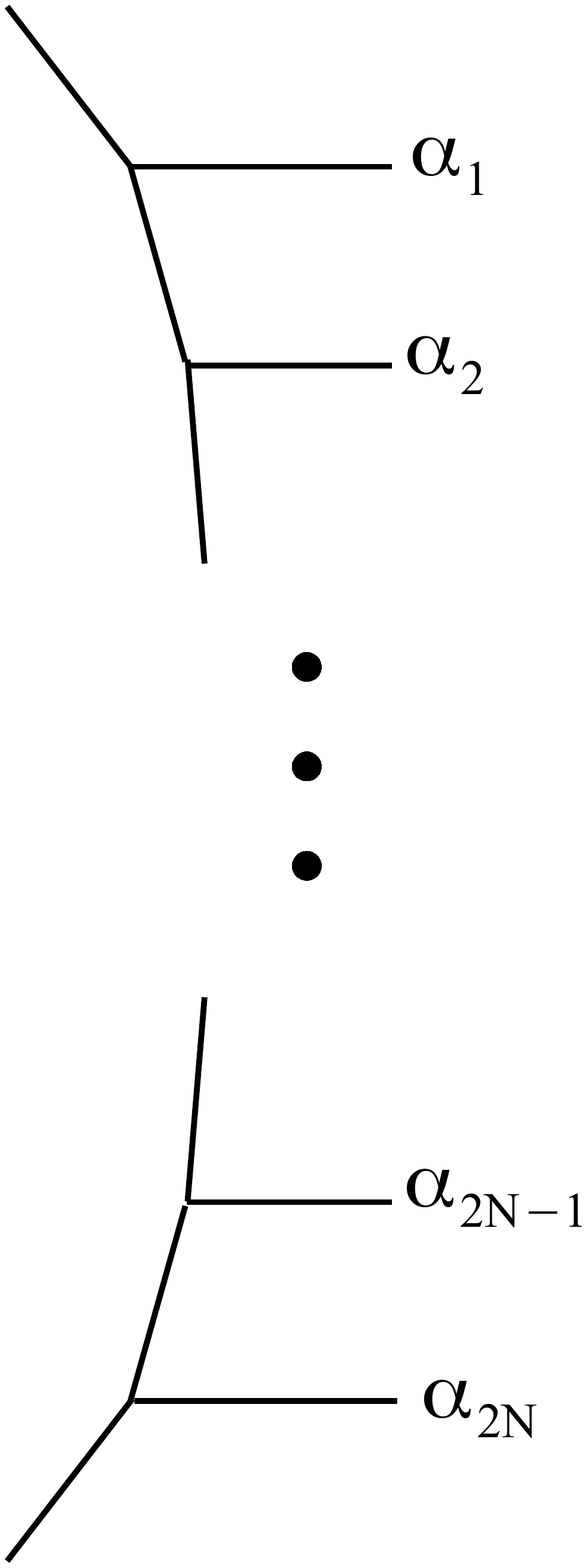}
\end{center}
\caption{The  ``half $SU(N)$ geometry''.}
\label{sun}
\end{figure}

We will now show how to geometrically  engineer  this theory.
The relevant geometry is now the  ``half $SU(N)$ geometry'' that is the resolved $A_{N-1}\times \mathbb{C}$ fibration
depicted in Fig. (\ref{sun}).
Actually, as explained in \cite{iq}, for fixed $N$, there are $N+1$ possible inequivalent geometries labelled by the integer $m=0,\cdots N$. Here we restrict to the $m=0$ case, however  other choices of $m$
correspond simply to the inclusion of extra framing factors.

The open  partition function  is given by \cite{iq}:

\ben
\label{css}
\frac{\mathcal{K}^{\alpha_1\cdots \alpha_{2N}} }{
\mathcal{K}^{\bullet \cdots \bullet}}=
\left(\prod_l^{2N} s_{\alpha_l}(q^\rho) \right)
\frac{1}{\prod_k\prod_{i< j}(1-q^k Q_i\cdots Q_{j-1})^{C_k(\alpha_i,\alpha_j^t)} }\, .
\een
We are interested in configurations involving a single toric brane placed in one of the $2N$ {\it gauge legs},
which correspond to take   $\alpha_i=1^n$ and $\alpha_j=0$ for $j\neq i$ in (\ref{css}).
Using the definition of the  $C_k$ coefficients given in eq. (\ref{ck}) we obtain:

\ben
A^{(i)}_n\equiv\frac{\mathcal{K}^{\bullet\cdots 1^n \cdots \bullet} }{
\mathcal{K}^{\bullet \cdots \bullet}}=
\frac{1}{\prod_{k=1}^n (1-q^k)}
\frac{   \left( q^{n(n-1)/2 } (-1)^{n}\right)^{(i-1)}  \left(Q_1 Q_2^2 Q_3^3\cdots Q_{i-1}^{(i-1)}\right)^{-1} }{\prod_{k=1}^n\prod_{ j>i}(1-q^{(k-1)} Q_i\cdots Q_{j-1})   \prod_{ j<i}(1-q^{(k-1)}  (Q_j\cdots Q_{i-1})^{-1})         } \, .\nonumber \\
\een

We now introduce a framing factor $\left( q^{n(n-1)/2 } (-1)^{n}\right)^{-(i-1)} $ for the toric brane placed in the
$i-$the leg and construct the generating function:
\be
\label{gen2}
Z^{(i)}_{top}=\sum_n A^{(i)}_n z^n\, ,
\ee
where we also included in the  open modulus the factor  $\left(Q_1 Q_2^2 Q_3^3\cdots Q_{i-1}\right)^{(i-1)}$.
It is now easy to see that the topological string partition function (\ref{gen2}) 
matches the vortex partition function  (\ref{cvo})  
with the following identifications:
\ben
 Q_i\cdots Q_{j-1}=q e^{-2\pi i b E_{ji}}\, ,\qquad   \left(Q_j\cdots Q_{i-1}\right)^{-1}=q e^{-2\pi i b E_{ij}}\, .
\een

 \section{Discussion}\label{s3}

We conclude with few comments about the generality of the factorisation.

As we explained in section \ref{s1} partition functions of abelian theories with arbitrary  number of chirals
and a  Chern-Simons term can  be expressed in a factorised form.
Indeed we observed that the condition ensuring  factorisation, being equivalent to the
cancelation of the $\mathbb{Z}_2$ anomaly, is  always satisfied.

Non abelian theories, such as the $U(N)$ theory with $N_f$ flavours can also be argued to factorise.
The strategy is to proceed  as in \cite{bp} and `abelianise' the integral by means of repeated applications of 
the Cauchy formula so to reduce the non-abelian partition function  to a sum of abelian ones.
It should be possible to interpret the final expression as a counting
of more general  vortex configurations labelled by Young tableaux.

However rather than proving factorisation in  a case by case analysis,
it would be more illuminating to derive it directly from an alternative localisation of the path integral.

In this note we did not discuss possible applications of our results 
in the context of the  recently proposed  correspondence \cite{dgl,DG,DGG,yama,ccv}
relating 3d  $\mathcal{N}=2$  gauge theories to analytically continued Chern-Simons theory.
This topic will be addressed elsewhere \cite{prog}.

\section*{Acknowledgments}
We would like to thank S.~Benvenuti, S.~Cremonesi,  C.~Kozcaz and T.~Dimofte for  discussions 
and in particular G.~Bonelli,  N.~Drukker and F.~Passerini
for discussions and   comments on the draft.
The work of the author is partially supported by a  Marie Curie Intra-European Fellowship: FP7-PEOPLE-2009-IEF.

\appendix

\section{Double-sine identities}\label{appendix}
The double-sine function is defined as:
\be
s_b(x)=\prod_{m,n\geq 0} \frac{m b +n/b + Q/2-i x}{m b +n/b + Q/2+i x}\, ,
\ee
and satisfies the following identities:
\ben
\label{s2c}
s_b(x)s_b(-x)=1\, , \qquad s_b(ib/2-x) s_b(ib/2+ x)=\frac{1}{2 \cosh\pi b x}\, ,
\een
or equivalently:
\be
s_b(iQ/2+x)=\frac{s_b(iQ/2+x-i b)}{ 2i \sinh\pi b x}\, .
\ee
From the above identities we deduce the following 
expressions useful to evaluate resides:
\ben
\label{master}
\nonumber \frac{s_b(x+iQ/2+ i m b +i n/b)}{s_b(x+ i Q/2)}&=&\frac{(-1)^{m n}}{\prod_{k=1}^n 2 i \sinh\pi b (x+i k b) \prod_{l=1}^m  2 i \sinh\pi/b(x+ i l/b)}=\\
&&=\frac{(-1)^{n m} (-i)^{n+m} q_1^{-n(n+1)/2} q_2^{-m(m+1)/2 }  e^{- \pi b x n} e^{-\pi/b x m} }{ \prod_{k=1}^n (1-q_1^{-2k} e^{- 2\pi b x})   \prod_{l=1}^m (1-q_2^{-2l} e^{-2\pi  x/b})}\, ,\nonumber\\
\een

\ben
\label{master2}
\nonumber\frac{s_b(x-iQ/2+ i m b +i n/b)}{s_b(x-iQ/2)}&=&\frac{(-1)^{m n}}{\prod_{k=1}^n 2 i \sinh\pi b (x-iQ+i k b) \prod_{l=1}^m  2 i \sinh\pi/b(x-i Q+ i l/b)}=\\&&=
\frac{(-1)^{n m} (-i)^{n+m} q_1^{-n(n+1)/2} q_2^{-m(m+1)/2 }  e^{-\pi b (x-iQ) n} e^{-\pi/b (x-iQ) m} }{ \prod_{k=1}^n (1-q_1^{-2(k-1)} e^{-2\pi b x})   \prod_{l=1}^m (1-q_2^{-2(l-1)} e^{-2\pi   x/b})}\, ,\nonumber\\
\een
where we defined:
 \ben
 q_1=e^{i \pi b^2}, \qquad q_2=e^{i \pi/ b^2}\, .
 \een

A useful factorised expression for the double-sine function:
\ben
\label{ss}
s_b(z)&=&e^{-i \pi z^2/2}
\frac{\prod_{k=1}^\infty\left(1+ e^{2\pi b z}e^{2\pi i b^2(k-1/2)}\right)}{\prod_{k=1}^\infty\left(1+e^{2\pi z/b}e^{2\pi i/ b^2(1/2-k)}\right)}=\nonumber \\&=&\nonumber
\frac{e^{-i \pi/2 z^2}}{\prod_{k=1}^\infty\left(1+q_1^{-2k} q_1 e^{2\pi bz)}\right)\prod_{k=1}^\infty\left(1+q_2^{-2k} q_2 e^{2\pi/bz}\right)}=\\ &=&
\frac{e^{-i \pi/2 z^2}}{\prod_{k=0}^\infty\left(1+q_1^{-(2k+1)}  e^{2\pi bz)}\right)\prod_{k=0}^\infty\left(1+q_2^{-(2k+1)} e^{2\pi/bz}\right)}\, ,
\een
and for its log:
\ben
\label{free}
\log s_b(z)&=&\frac{-i \pi z^2}{2}
+\left( \sum_k \frac{(-1)^k}{k} \frac{ e^{2\pi  z b k}  }{(q_1^k-q_1^{-k})}\right)+
\left( \sum_k \frac{(-1)^k}{k}  \frac{ e^{2\pi  z/ b k}  }{(q_2^k-q_2^{-k})}\right)\, ,
\een
can be derived by using the following analytic continuation:
\ben
\label{ana}
\prod_{n=0}^\infty (1+ q^{-(2n+1)} w)=\exp\left(-\sum_{k=1} \frac{(-1)^k w^k}{k(q^k-q^{-k})} \right)=1/\prod_{n=0}^\infty (1+ q^{2n+1} w)\,.
\een

We also need the  $b\to 0$ limit:

\be
\label{limit}
s_b(z)\to e^{-\pi i z^2/2} e^{i \pi(2-Q^2)/24} \exp\left(\frac{1}{2\pi i b^2} {\rm Li}_2(-e^{2\pi b z})\right)\, .
\ee

%%%%%%%%%%%%%%%%%%%%%%%%%%%%%%%%%%%%%%%%%%%%%%%%%%%%%%%%%%%%%%%%%

%%%%%%%%%%%%%%%%%%%%%%%%%%%%%%%%%%%%%%%%%%%%%%%%%%%%%%%%%%%%%%%%%


\begin{thebibliography}{99}
\bibliographystyle{plain}

\bibitem{hama}
  N.~Hama, K.~Hosomichi and S.~Lee,
  ``SUSY Gauge Theories on Squashed Three-Spheres,''
  JHEP {\bf 1105} (2011) 014
  [arXiv:1102.4716 [hep-th]].
  %%CITATION = JHEPA,1105,014;%%



\bibitem{pestun}
  V.~Pestun,
  ``Localization of gauge theory on a four-sphere and supersymmetric Wilson
  loops,''
  {\tt arXiv:0712.2824}.
  %%CITATION = ARXIV:0712.2824;%%



%\cite{Kapustin:2009kz}
\bibitem{kapo}
  A.~Kapustin, B.~Willett and I.~Yaakov,
  ``Exact Results for Wilson Loops in Superconformal Chern-Simons Theories with
  Matter,''
  JHEP {\bf 1003} (2010) 089
  [arXiv:0909.4559 [hep-th]].
  %%CITATION = JHEPA,1003,089;%%
%\cite{Hama:2011ea}








\bibitem{dgl}
  T.~Dimofte, S.~Gukov, L.~Hollands,
  ``Vortex Counting and Lagrangian 3-manifolds,''
  [arXiv:1006.0977 [hep-th]].


%\cite{Dimofte:2011jd}
\bibitem{DG}
  T.~Dimofte, S.~Gukov,
  ``Chern-Simons Theory and S-duality,''
    [arXiv:1106.4550 [hep-th]].




%\cite{Dimofte:2011ju}
\bibitem{DGG}
  T.~Dimofte, D.~Gaiotto, S.~Gukov,
  ``Gauge Theories Labelled by Three-Manifolds,''
    [arXiv:1108.4389 [hep-th]].



%\cite{Shadchin:2006yz}
\bibitem{shad}
  S.~Shadchin,
  ``On F-term contribution to effective action,''
  JHEP {\bf 0708} (2007) 052
  [arXiv:hep-th/0611278].
  %%CITATION = JHEPA,0708,052;%%





%\cite{Eynard:2008he}
\bibitem{nonp}
  B.~Eynard and M.~Marino,
  ``A holomorphic and background independent partition function for matrix
  models and topological strings,''
  J.\ Geom.\ Phys.\  {\bf 61} (2011) 1181
  [arXiv:0810.4273 [hep-th]].
  %%CITATION = JGPHE,61,1181;%%




%\cite{Bonelli:2011fq}
\bibitem{btz}
  G.~Bonelli, A.~Tanzini, J.~Zhao,
  ``Vertices, Vortices and Interacting Surface Operators,''
    [arXiv:1102.0184 [hep-th]]


%\cite{Ooguri:2004zv}
\bibitem{osv}
  H.~Ooguri, A.~Strominger and C.~Vafa,
  ``Black hole attractors and the topological string,''
  Phys.\ Rev.\  D {\bf 70} (2004) 106007
  [arXiv:hep-th/0405146].
  %%CITATION = PHRVA,D70,106007;%%




\bibitem{sv}
  V.~P.~Spiridonov and G.~S.~Vartanov,
  ``Elliptic hypergeometry of supersymmetric dualities II. Orthogonal groups, knots, and vortices,''
  arXiv:1107.5788 [hep-th].
  %%CITATION = ARXIV:1107.5788;%%



\bibitem{hv}
  K.~Hori and C.~Vafa,
  ``Mirror symmetry,''
  arXiv:hep-th/0002222.
  %%CITATION = HEP-TH/0002222;%%
  
  
\bibitem{AV}
  M.~Aganagic and C.~Vafa, 
``Mirror symmetry, D-branes and counting holomorphic discs,''
 {\tt  hep-th/0012041}.
  %%CITATION = HEP-TH 0012041;%%
 
\bibitem{AKV}
 M.~Aganagic, A.~Klemm and C.~Vafa,
``Disk instantons, mirror symmetry and the duality web,''
  {\em Z.\ Naturforsch.\ A} {\bf 57}, 1 (2002)
  {\tt hep-th/0105045}.
  %%CITATION = HEP-TH 0105045;%%





  
  %\cite{Bouchard:2007ys}
\bibitem{remo}
  V.~Bouchard, A.~Klemm, M.~Marino and S.~Pasquetti,
  ``Remodeling the B-model,''
  Commun.\ Math.\ Phys.\  {\bf 287} (2009) 117
  [arXiv:0709.1453 [hep-th]].
  %%CITATION = CMPHA,287,117;%%

%\cite{Kozcaz:2010af}
\bibitem{ab}
  C.~Kozcaz, S.~Pasquetti and N.~Wyllard,
  ``A \& B model approaches to surface operators and Toda theories,''
  JHEP {\bf 1008} (2010) 042
  [arXiv:1004.2025 [hep-th]].
  %%CITATION = JHEPA,1008,042;%%

  
  
  \bibitem{vertex}
  M.~Aganagic, A.~Klemm, M.~Mari\~no and C.~Vafa,
  ``The topological vertex,''
  {\em Commun.\ Math.\ Phys.\ }  {\bf 254} (2005) 425, 
  {\tt hep-th/0305132}.
  %%CITATION = CMPHA,254,425;%%




  

%\cite{Iqbal:2004ne}
\bibitem{strip}
  A.~Iqbal and A.~K.~Kashani-Poor,
  ``The vertex on a strip,''
  Adv.\ Theor.\ Math.\ Phys.\  {\bf 10} (2006) 317
  [arXiv:hep-th/0410174].
  %%CITATION = 00203,10,317;%%



%\cite{Tachikawa:2004ur}
\bibitem{tac}
  Y.~Tachikawa,
  ``Five-dimensional Chern-Simons terms and Nekrasov's instanton counting,''
  JHEP {\bf 0402} (2004) 050
  [arXiv:hep-th/0401184].
  %%CITATION = JHEPA,0402,050;%%


\bibitem{iq}
A.~Iqbal and A.-K. Kashani-Poor, ``{SU($N$) geometries and topological string
  amplitudes},'' {\em Adv. Theor. Math. Phys.} {\bf 10} (2006) 1--32,
{{\tt hep-th/0306032}}; \\







%\cite{Benvenuti:2011ga}
\bibitem{bp}
  S.~Benvenuti and S.~Pasquetti,
  ``3D-partition functions on the sphere: exact evaluation and mirror
  symmetry,''
  arXiv:1105.2551 [hep-th].
  %%CITATION = ARXIV:1105.2551;%%

\bibitem{yama}
  Y.~Terashima and M.~Yamazaki,
  ``SL(2,R) Chern-Simons, Liouville, and Gauge Theory on Duality Walls,''
  JHEP {\bf 1108} (2011) 135
  [arXiv:1103.5748 [hep-th]].
  %%CITATION = JHEPA,1108,135;%%


%\cite{Cecotti:2011iy}
\bibitem{ccv}
  S.~Cecotti, C.~Cordova and C.~Vafa,
  ``Braids, Walls, and Mirrors,''
  arXiv:1110.2115 [hep-th].
  %%CITATION = ARXIV:1110.2115;%%


%\cite{Benvenuti:2011ga}
\bibitem{prog}
Work in progress.
  %%CITATION = ARXIV:1105.2551;%%



\end{thebibliography}
\end{document}